\documentclass[%
showpacs,
 amsmath,amssymb,
 aps,
 twocolumn,
pre,
 balancelastpage
]{revtex4-1}

\usepackage{graphicx}
\usepackage{dcolumn}
\usepackage{bm}
\usepackage{hyperref}
\usepackage{amssymb}
\usepackage{amsmath}
\usepackage{epsfig}
\usepackage{wrapfig}
\usepackage{verbatim}
\usepackage{sidecap}
\usepackage{epstopdf}
\usepackage{color}

\newcommand{\Eq}[1]{(\ref{#1})}

\newcommand{\beq}[1]{\begin{equation}\label{#1}}
\newcommand{\eeq}{\end{equation}}

\newenvironment{se}[1]{\equation\label{#1}\aligned}{\endaligned\endequation}
\newcommand{\bsplit}[1]{\begin{se}{#1}}
\newcommand{\esplit}{\end{se}}

\def\Xint#1{\mathchoice
   {\XXint\displaystyle\textstyle{#1}}%
   {\XXint\textstyle\scriptstyle{#1}}%
   {\XXint\scriptstyle\scriptscriptstyle{#1}}%
   {\XXint\scriptscriptstyle\scriptscriptstyle{#1}}%
   \!\int}
\def\XXint#1#2#3{{\setbox0=\hbox{$#1{#2#3}{\int}$}
     \vcenter{\hbox{$#2#3$}}\kern-.5\wd0}}
\def\dashint{\Xint-}

\begin{document}

\preprint{APS/123-QED}

\title{Onset of Synchronization in the Disordered Hamiltonian Mean Field Model}

\author{Juan G. Restrepo}
\email{juanga@colorado.edu}
\author{James D. Meiss}
\email{James.Meiss@colorado.edu}
\affiliation{
Department of Applied Mathematics, University of Colorado at Boulder, Boulder, Colorado 80309, USA
}%

\date{\today}

\begin{abstract}
We study the Hamiltonian Mean Field (HMF) model of coupled Hamiltonian rotors with a heterogeneous distribution of moments of inertia and coupling strengths. We show that when the parameters of the rotors are heterogeneous, finite size fluctuations can greatly modify the coupling strength at which the incoherent state loses stability by inducing correlations between the momenta and parameters of the rotors. When the distribution of initial frequencies of the oscillators is sufficiently narrow, an analytical expression for the modification in critical coupling strength is obtained that confirms numerical simulations. We find that heterogeneity in the moments of inertia tends to stabilize the incoherent state, while heterogeneity in the coupling strengths tends to destabilize the incoherent state. Numerical simulations show that these effects disappear for a wide, bimodal frequency distribution. 
\end{abstract}

\maketitle

\section{Introduction}

Hamiltonian systems with long range interactions appear in many areas of physics, including systems with gravitational or Coulomb interactions \cite{chavanis2005dynamics, tatekawa2005thermodynamics}, vortex dynamics in fluids \cite{Aref83, weiss1998lagrangian, chavanis2002statistical}, plasma physics \cite{Tennyson94, CastilloNegrete02, Balmforth12}, and free-electron lasers \cite{Antoniazzi06, bachelard2010experimental}. Many of the generic properties of emergent collective behavior in these systems can be studied with the Hamiltonian Mean Field (HMF) model \cite{konishi1992clustered, inagaki1993dynamical, antoni1995clustering}. In this model, $N$ inertial rotors 
described by their conjugate phase, $\theta_n$, and angular momentum, $p_n$, variables interact under the Hamiltonian
\begin{equation}\label{HMF}
H = \sum_{n=1}^N \frac{p_n^2}{2I_n} - \frac{1}{2}\frac{K}{N} \sum_{n,m=1}^N a_n a_m \cos(\theta_m - \theta_n),  
\end{equation}
where the $n^{th}$ rotor has moment of inertia $I_n$ and coupling constant $a_n$. One can also view \Eq{HMF} as describing an interacting set of particles on a periodic domain (the circle) with \textit{masses} $I_n$, and  \textit{momenta} $p_n$, that are coupled through \textit{charges} $a_n$. Despite its simplicity and analytical tractability, the HMF model exhibits some of the features present in many Hamiltonian systems with long-range interactions including violent relaxation towards long lived quasistationary states, slow collisional relaxation, and phase transitions  \cite{chavanis2005dynamics, chavanis2002statistical}. Because of these similarities, the HMF model has become an iconic testbed for the study of Hamiltonian systems with long range interactions  \cite{barre2006vlasov, campa2007long, dauxois2002hamiltonian, latora1999chaos, yamaguchi2004stability}.
Previous studies (with the exception of \cite{chavanis2005dynamics, ciani2011long, de2013critical})  have assumed that the rotors have identical moments of inertia and that they are equally coupled to all other rotors; i.e., $I_n = a_n = 1$. However, in many physical systems there is often some form of disorder. For example, stars in self-gravitating systems have a heterogeneous mass distribution \cite{chavanis2005dynamics}, and vortices in 2D turbulence have a heterogeneous circulation distribution \cite{bracco2000revisiting}. We study the effects of disorder in the HMF model to provide insights that may be valid more generally. The model \Eq{HMF} allows for the study of rotors with different masses or lengths (variable $I_n$), and disorder in the interactions due to gravity (mass $a_n \propto I_n$) or Coulomb forces (charge $\propto a_n$). 
We will assume that $I_n$ and $a_n$ are chosen from a joint distribution $h(I,a)$, and that $h$ vanishes when $I$ or $a$ is negative; the case that $a >0$ can also be thought of as \textit{ferromagnetic}. Since the coupling strength is represented by the parameter $K$, without loss of generality, we can assume that the mean coupling strength is one: $\langle a \rangle = 1$, where $\langle \cdot \rangle$ denotes an average over the distribution $h$. We will show that even a small heterogeneity in the distribution of these parameters can drastically affect the onset of synchronization. 

This paper is structured as follows. In Section~\ref{theory}, we present our theory and derive an expression for the critical coupling strength  at the onset of synchronization as a function of the distribution of the parameters $I$ and $a$ as well as of the distribution of initial momenta. In Section~\ref{numerics} we illustrate our results with numerical experiments. We present our conclusions in Sec.~\ref{conclusion}.

\section{Effect of Disorder on the Onset of Instability}\label{theory}

We are interested in  the effect of disorder on the transition to synchronized behavior in the system described by Hamiltonian (\ref{HMF}). The canonical equations corresponding to \Eq{HMF} are
\beq{can}
	(\dot \theta_n, \dot p_n) = \left( \frac{p_n}{I_n}, -K a_n R \sin(\theta_n-\psi)\right),
\eeq

where the complex order parameter is defined as
\beq{orderParam}
	R e^{i\psi} \equiv \frac{1}{N}\sum_{n=1}^N a_n e^{i \theta_n} ;
\eeq
analogous to that used in the Kuramoto model \cite{acebron2005kuramoto, antoni1995clustering}.

In the continuum limit, $N \to \infty$, this system can be formulated in terms of the density $\rho(\theta,p,t;I,a)$ of rotors with phase $\theta$, momentum $p$, mass $I$, and charge $a$ at time $t$. In this limit, the order parameter becomes 
\[
	R e^{i\psi} =  
	\int_0^\infty\int_0^\infty \int_{-\infty}^\infty \int_0^{2\pi} 
	    a e^{i\theta} \rho(\theta,p,t;I,a)d\theta d p d I da.
\] 
The evolution of the density is given by the continuity equation (in this context often referred to as a Vlasov equation \cite{bouchet2005prediction, barre2006vlasov})
\beq{liouville}
	\frac{\partial \rho}{\partial t} + \frac{p}{I} \frac{\partial \rho}{\partial \theta} - 
	    aKR\sin(\theta-\psi) \frac{\partial \rho}{\partial p} = 0.
\eeq
This equation admits {\it incoherent} equilibria corresponding to densities of the form $\rho = G(p; I, a)/(2\pi)$, for which  $R = 0$. Since the masses and charges do not evolve in time, it is convenient isolate their distribution, $h(I,a)$, and rewrite $G(p;I,a) = g_{I,a}(p) h(I,a)$; thus $g_{I,a}(p)$ is the distribution of momenta conditioned on $I$ and $a$. 

A linear stability analysis of the incoherent solution, analogous to that in \cite{strogatz1991stability, inagaki1993dynamical, choi2003stability}, implies the existence of a critical coupling strength $K_c$ and critical frequency $\omega$ for the onset of instability, determined by
\bsplit{oonset}
	1  &=  -\frac{K_c}{2 }\int_0^{\infty}\int_0^{\infty} a^2 I h(I,a) 
			\dashint_{-\infty}^{\infty}\frac{g_{I,a}'(p)}{p - I \omega}dp \,d I da,\\
	0  &=  \int_0^{\infty}\int_0^{\infty} a^2 I h(I,a) g_{I,a}'(I \omega) dI da,
\esplit
where $g_{I,a}'(p) = \partial g_{I,a}(p)/\partial p$ and $\dashint$ denotes the principal value integral. 
From this analysis, one would expect that if the initial density $G$ is known and $K$ is increased adiabatically, the rotors would remain incoherent for $K\leq K_c$ (i.e., $R\approx 0$), and that the incoherent state would lose its stability for $K>K_c$ (i.e., $R$ would become nonzero).
By contrast, numerical simulations with the simple incoherent initial distribution
$
	\rho(\theta,p,0;I,a) =  g_0(p)h(I,a)/(2\pi),
$
for which $g_0$ is independent of the parameters, show that this state may remain stable for values of $K$ much larger than the predicted $K_c$ (e.g., as we will show later, for a narrow distribution of masses with a relative width $\sim 0.1$, the system can remain stable up to $K \sim 10 K_c$). Our goal is to understand this modification to the onset of instability. We will obtain analytical results under the assumption that the distribution of effective frequencies $\dot \theta = p/I$ [cf.~\Eq{can}] has a ``small enough" width around its mean. Simulations will show what happens when this assumption is violated.

To begin, we note that the linear stability analysis that leads to the dispersion relation \Eq{oonset} also shows that perturbations to the density that are independent of $\theta$ are marginally stable. In the classical analysis of Strogatz and Mirollo for the Kuramoto model, these perturbations are forbidden in order to conserve the number of oscillators \cite{strogatz1991stability}. For \Eq{liouville}, however, any incoherent perturbation $\rho \to G(p;I,a)/(2\pi) + \eta(p;I,a)$  with zero average
is allowed and undamped. Such fluctuations should be expected in the simulations for finitely many oscillators, and these may modify the predicted threshold from \Eq{oonset}.

To quantify this we assume that, instead of being zero, the order parameter \Eq{orderParam} fluctuates around a small rotating term, i.e., $R e^{i\psi}  = \bar R e^{i\Omega t} + z$, where $\bar R \ll 1$,  $\Omega$ is a coherent frequency that will be determined self-consistently, and $z$ represents time dependent fluctuations with zero mean.   For the Kuramoto model it has been shown that $\bar R \sim N^{-1/2}$ \cite{hildebrand2007kinetic}, and numerical simulations \cite{yamaguchi2004stability} have shown the same scaling holds for the HMF model. 
The existence of a dominant frequency $\Omega$ should be reasonable if the initial distribution of frequencies is narrow enough. Defining $\tilde \theta_n =\theta_n-\Omega t$ and $\tilde p_n = p_n - I_n \Omega$, \Eq{can} becomes
\[
	\dot{\tilde \theta}_n = \frac{\tilde p_n}{I_n}, \quad
	\dot{\tilde p}_n = -Ka_n\bar R \sin(\tilde \theta_n) -Ka_n\text{Im}(z e^{-i \theta_n}).
\]
In the absence of the fluctuating term, the rotors are decoupled and each has constant energy $\bar E_n = \tilde p_n^2/(2 I_n) - K a_n \bar R \cos(\tilde \theta_n)$.~We treat the term $Ka_n\text{Im}(z e^{-i \theta_n})$ as a stochastic perturbation to the Hamiltonian dynamics, with the important characteristic that this perturbation conserves the total energy \Eq{HMF}. In the context of the HMF model, it has been shown that the stationary distribution of energies in the ensemble is numerically very close to a Boltzmann distribution \cite{yamaguchi2004stability}, i.e., the density of rotors with energy $\bar E$ is proportional to $\exp(-\bar E/\sigma^2)$, where $\sigma^2$ is the temperature.  Thus below the onset of synchronization, letting $\bar R = 0$, the stationary density of rotors with momentum $p$, given $I$ and $a$, becomes
\beq{finaldist}
	g_{I,a} (p) = \frac{1}{\sqrt{2\pi I \sigma^2}} 
		\exp\left(-\frac{\left(p - I \Omega \right)^2}{2  I \sigma^2}\right),
\eeq
a Gaussian distribution with mean $I \Omega$ and variance $I \sigma^2$. This density will evolve from the initial density, $g_0(p)$, over a time scale to be determined.

To determine $\Omega$, we note that both the discrete \Eq{can} and continuum \Eq{liouville} systems preserve the average momentum $\langle p \rangle$ which is initially $P = \int_{-\infty}^{\infty}pg_0(p)dp$ and becomes $\int_0^{\infty}\int_0^{\infty}\int_{-\infty}^{\infty} p g_{I,a}(p)h(I,a) dp dI da= \langle I \rangle \Omega$ in the stationary state. Therefore 
\beq{shift}
	\Omega = P \langle I \rangle^{-1}.
\eeq
Similarly, conservation of energy determines the temperature $\sigma^2$ in terms of the variance $\sigma_0^2$ of the original distribution of momenta $g_0(p)$. 
Equating the initial energy and the energy in the stationary state with distribution \Eq{finaldist} gives
\[
	\sigma^2
	=  \langle I^{-1} \rangle \left[\sigma_0^2  +	
	    P^2\left(1  - \frac{1}{\langle I \rangle \langle I^{-1}\rangle}\right) \right]. 
\]

We now proceed to calculate the critical coupling strength from \Eq{oonset} for the marginal distribution \Eq{finaldist}. Since $g'_{I,a}(I\Omega) = 0$, then $\omega = \Omega$. Inserting this in the first equation of \Eq{oonset} gives our main result,
\beq{kcfin}
K_c 
      \approx 2 \frac{\langle I^{-1} \rangle}{\langle a^2\rangle}\left[\sigma_0^2 +
		P^2\left(1 - \frac{1}{\langle I \rangle \langle I^{-1} \rangle}\right)\right].
\eeq

To compare this result with the homogeneous HMF model, consider the case $h(I,a) = \delta(I-I_0)\delta(a-a_0)$. If the initial distribution $g_0(p)$ were itself Gaussian, then \Eq{kcfin} predicts
\[
	K_c^0 = 2\frac{\sigma_0^2}{a_0^2 I_0} . 
\]
First let us consider the effect of heterogeneity in the distribution of masses. Since the Cauchy-Schwarz inequality implies $\langle I \rangle \langle I^{-1} \rangle \geq 1$, we have $K_c \geq K_c^0$. Thus mass heterogeneity enhances the stability of the incoherent state. 

On the other hand, whenever there is heterogeneity in the charge distribution, the factor $\langle a^2 \rangle$ in \Eq{kcfin} will be larger than $a_0^2$. Thus charge heterogeneity reduces $K_c$, destabilizing the incoherent state. We also note that if the initial distribution $g_0(p)$ is not Gaussian, \Eq{kcfin} might predict a lower $K_c$ than what would have been obtained using $g_0(p)$ in \Eq{oonset}.

\section{Numerical Experiments}\label{numerics}

In this Section, we will illustrate our results from Sec.~\ref{theory} with three numerical experiments. In example (i) the initial distribution $g_0(p)$ is a Gaussian centered at $P$ with standard deviation $\sigma_0 = 0.35$. There are $N = 1000$ rotors with a uniform distribution of $I$ on the interval $[1-\varepsilon,1+\varepsilon]$, but all have $a = 1$. In this case, it can be shown that the coefficient of variation (standard deviation over mean) of the distribution of the effective frequencies is bounded by $\sqrt{2(\sigma_0/P)^2 + \varepsilon^2}$ for $\varepsilon < 0.8$, so we expect our theory to apply when $\varepsilon \ll 1$ and $P \gg \sigma_0 = 0.35$.

\begin{figure}[b]
  \centering
    \includegraphics[width=0.95\linewidth]{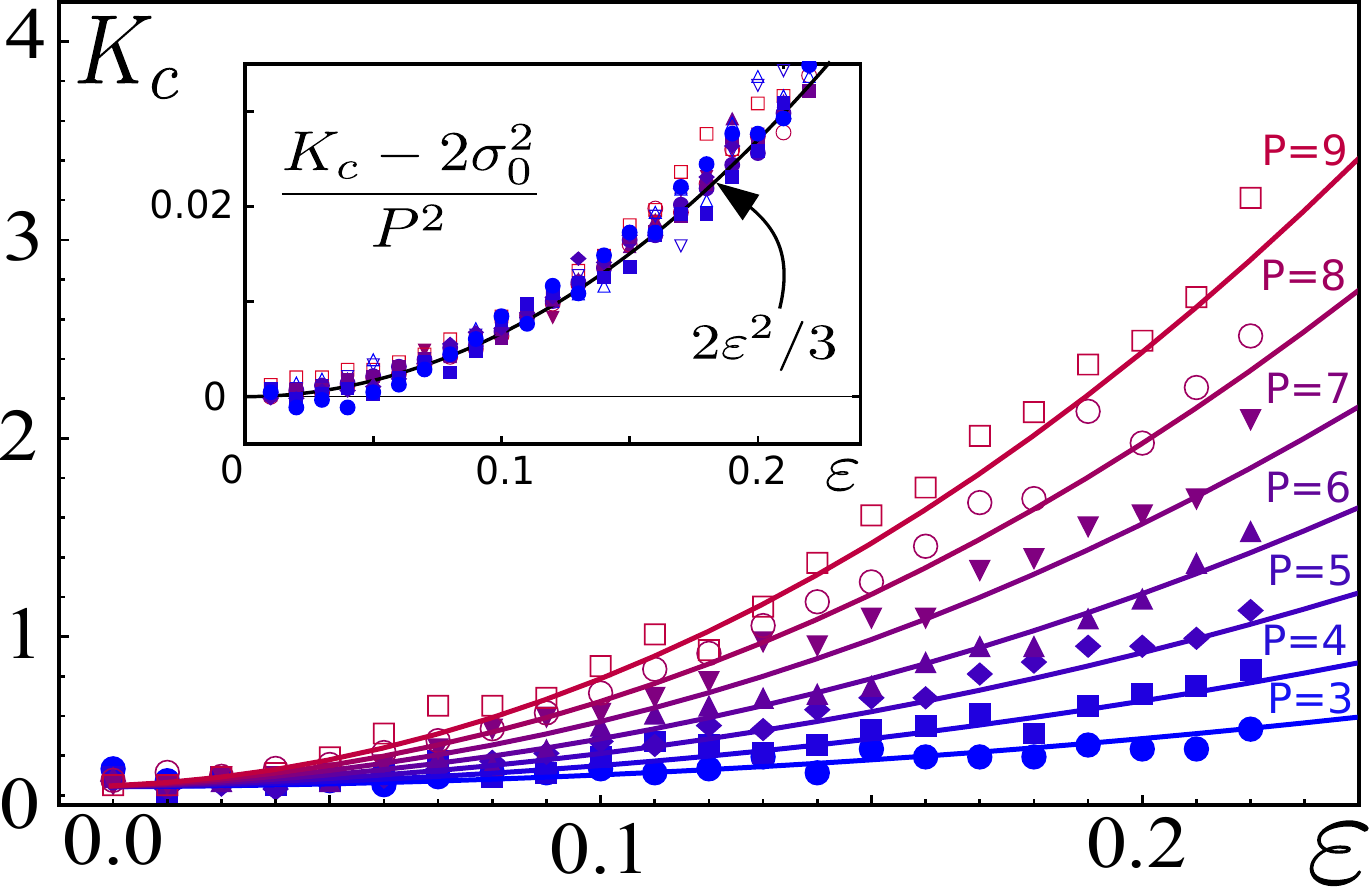}
    \caption{Estimated critical coupling as a function of width of the mass distribution, $\varepsilon$, for example (i) using average momenta as indicated (symbols), and corresponding $K_c$ predicted from \Eq{kcfin} (solid lines). Inset: estimated $(K_c - 2\sigma_0^2)/P^2$ compared to the prediction $2 \varepsilon^2/3$. The inset also includes simulations obtained using an uniform initial distribution of momenta (see text).}\label{kacs}
\end{figure}

First we test our prediction \Eq{kcfin} for $K_c$. For a given value of $P$ and $\varepsilon$, we simulate \Eq{can} by increasing $K$ by $0.02$ every $10,000$ time units from the initial value $K = 0$, and use the last $5000$ time units for each $K$ to estimate $R$ by a time average. The instability threshold was estimated as the value of $K$ at which the averaged $R$ last exceeds $0.1$. While this a rough estimate intended to be used for relatively small $N$, it allows us to see how $K_c$ varies as the parameters of the system are changed. In Fig.~\ref{kacs}(a) we plot the estimated $K_c$ as a function of the width $\varepsilon$ for the seven values of average momentum $P$ shown. The curves show the prediction
\[
	K_c(\varepsilon) = \frac{2}{\varepsilon}\mbox{atanh}(\varepsilon)
	    \left[\sigma_0^2 + P^2\left(1- \frac{\varepsilon}{\mbox{atanh}(\varepsilon)}\right)
	         \right]
\]
of \Eq{kcfin}. The inset shows that the computed value of $(K_c - 2\sigma_0^2)/P^2$ collapses onto the solid black curve, given from the theory (for small $\varepsilon$ and $\sigma_0$) by $2 \varepsilon^2/3$. Since our results depend only on the mean $P$ and variance $\sigma_0$ of $g_0(p)$, we also consider a uniform distribution of initial momenta $g_0(p)$ with the same standard deviation $\sigma = 0.35$ and various $P$. The rescaled values of $K_c$ are also included in the inset of Fig.~\ref{kacs}, and they collapse onto the same curve.

\begin{figure}[t]
\setlength{\belowcaptionskip}{-18pt}
  \centering
    \includegraphics[width=\linewidth]{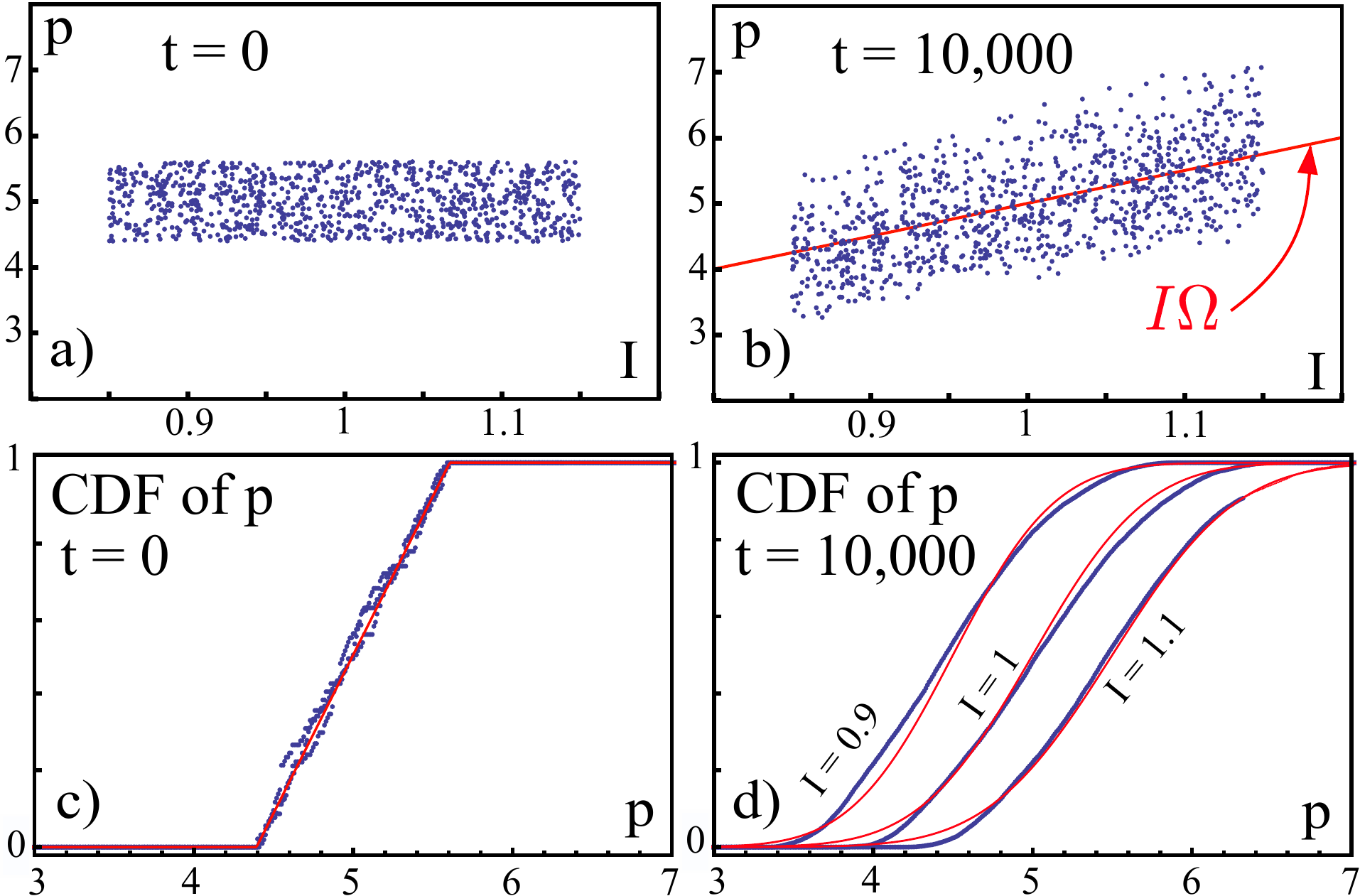}
    \caption{(Color online) Evolution of the momentum-mass distribution for $1000$ rotors each with $a=1$, given a uniform distribution of masses $I$ with $\langle I \rangle = 1$ and of momenta with $\langle p \rangle = P =5$. (a) Momentum-mass distribution at $t=0$ with half-width $\varepsilon = 0.15$ in mass and standard deviation $\sigma_0 = 0.35$ in momentum, and (b) for $t = 10,000$ when $K=0.2$. The solid line is $ \langle p \rangle =  I \Omega$ with $\Omega = P = 5$; (c) CDFs of momenta $p$ conditioned on $I$  for $I = 0.9$, $1$, and $1.1$, for $t=0$, and (d) for $t = 10,000$. Theoretical CDFs are also shown (thin red curves).}\label{distributions}
\end{figure}

To illustrate the validity of \Eq{finaldist} in the incoherent state, we plot, in Fig.~\ref{distributions}(a), the initial masses and momenta of $N = 1000$ rotors using a uniform distribution of initial momenta $g_0(p)$, and, in Fig.~\ref{distributions}(b), their masses and momenta at $T=10,000$ for $K/K_c = 0.32$.
The solid line in Fig.~\ref{distributions}(b) indicates the predicted mean $I \Omega$ of the stationary distribution $g_{I,a}(p)$ using \Eq{shift}. The empirical cumulative distribution function (CDF) of momenta $p$ conditioned on $I$ is shown in Fig.~\ref{distributions}(c) at $t=0$, for three values of $I$. These were calculated from the rotors with masses in three slices $(I -0.025,I + 0.025)$ of the distribution. The curves are simply the theoretical CDF for the uniform distribution. In Fig.~\ref{distributions}(d) we show the CDFs at $t=10,000$ for the same three values of $I$ averaged over the time interval $[2500,10000]$. The curves show that the corresponding theoretical Gaussian CDFs, calculated using \Eq{finaldist}, agree reasonably well with the numerical distributions.

\begin{figure}[htb]
  \centering
    \includegraphics[width=\linewidth]{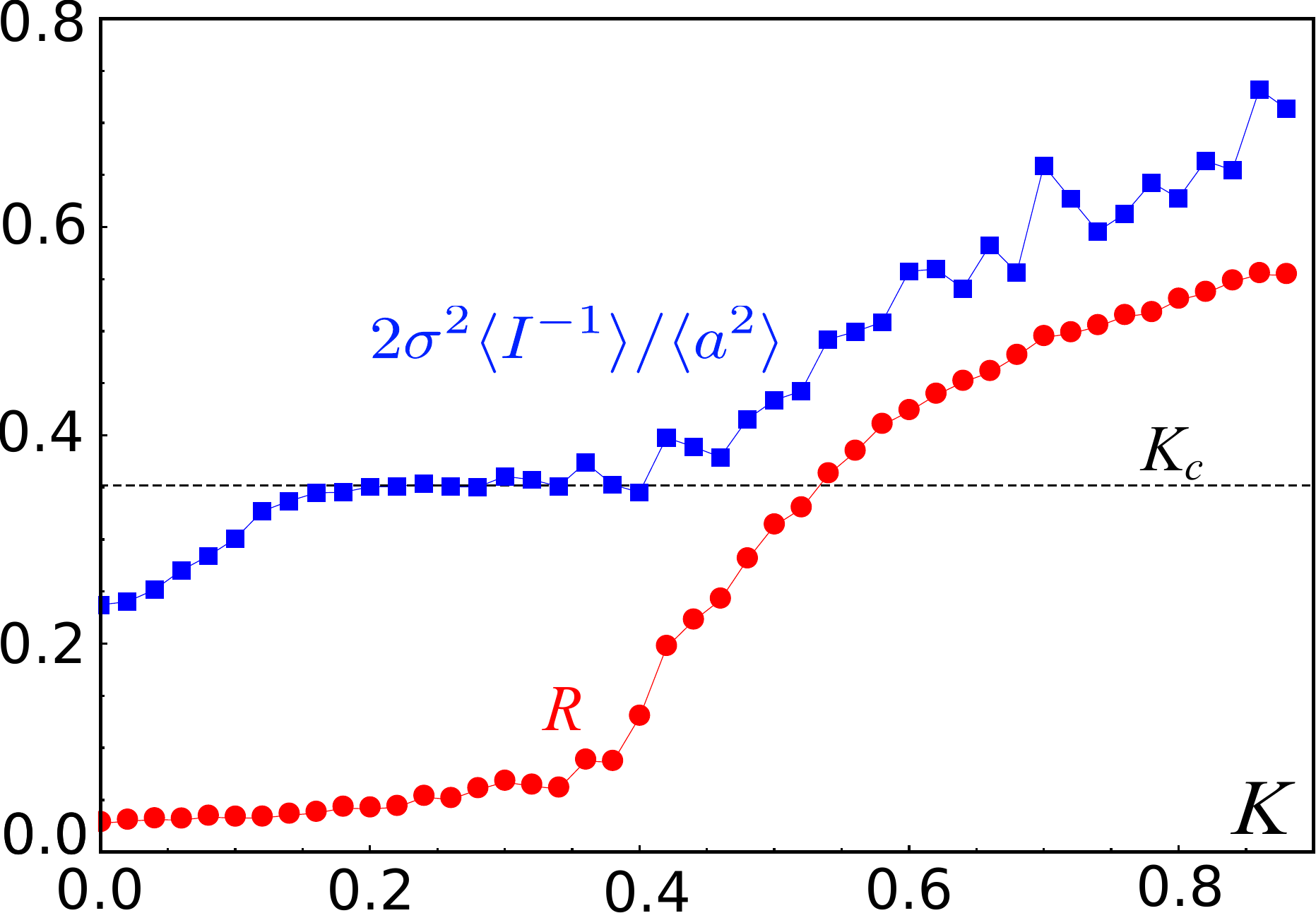}
    \caption{Order parameter $R$ (circles) and estimated $2\sigma^2 \langle I^{-1}  \rangle/\langle a^2  \rangle$ (squares) as $K$ is increased for $\varepsilon = 0.1$ and $P=5$ in example (i). The solid line indicates the stationary value of $K_c$ predicted by \Eq{kcfin}.}\label{times}
\end{figure}

Our analysis assumes that the initial distribution $g_0(p)$ relaxes to the stationary distribution $g_{I,a}(p)$; that is, the system must remain incoherent long enough for this relaxation to occur. To illustrate this, we estimate $\sigma^2$ from the simulations as the mean of the instantaneous conditional variance, $\langle\sigma_{I,a}^2\rangle$. According to \Eq{kcfin}, the effective critical coupling strength $2\sigma^2 \langle I^{-1} \rangle/\langle a^2  \rangle$ should approach $K_c$ as the distribution relaxes to Boltzmann. This quantity is shown in Fig.~(\ref{times}) (squares), as a function of $K$ when it is increased, again by $0.02$ each $t=10,000$ time units. Initially, the effective critical coupling strength is $2\sigma_0^2$, but as $K$ grows adiabatically, the distribution relaxes to \Eq{finaldist} and the effective critical coupling strength approaches the value predicted by \Eq{kcfin} (dashed line). 
When $K$ becomes larger than this critical value, the incoherent state loses its stability, as can be seen in the evolution of the order parameter $R$ (circles in the figure). If $K$ is increased too rapidly or $N$ is too large, the relaxation of the initial distribution to its steady state might not occur and the incoherent state could become unstable earlier. Indeed, in experiments with a uniform initial distribution (not shown in the figure) the relaxation is slower than when it is a Gaussian, and so $K$ must be increased more slowly to allow the distribution to relax to its steady state.

In agreement with previous observations \cite{latora1999chaos}, we observe that the relaxation time scales linearly with $N$. Thus our results apply in situations in which $N$ is not too large or when the system is allowed to reach equilibrium (e.g., as when we slowly changed $K$). In addition, if \Eq{kcfin} predicts a value smaller than $\hat K_c$ predicted for the initial distribution, $g_0(p)$, our computations show that when $K \in [K_c, \hat K_c]$ the incoherent state is only metastable, since it will become unstable as $g_0(p)$ relaxes to (\ref{finaldist}) \cite{pluchino2004metastable}.

In conclusion, example (i) has shown how heterogeneity in the masses combined with nonzero total momentum $P$ results in the stabilization of the incoherent state: $K_c$ is increased compared with the case of identical oscillators. Heterogeneity in the distribution of charges $a$ has the opposite effect. 

\begin{figure}[bht]
  \centering
    \includegraphics[width=1\linewidth]{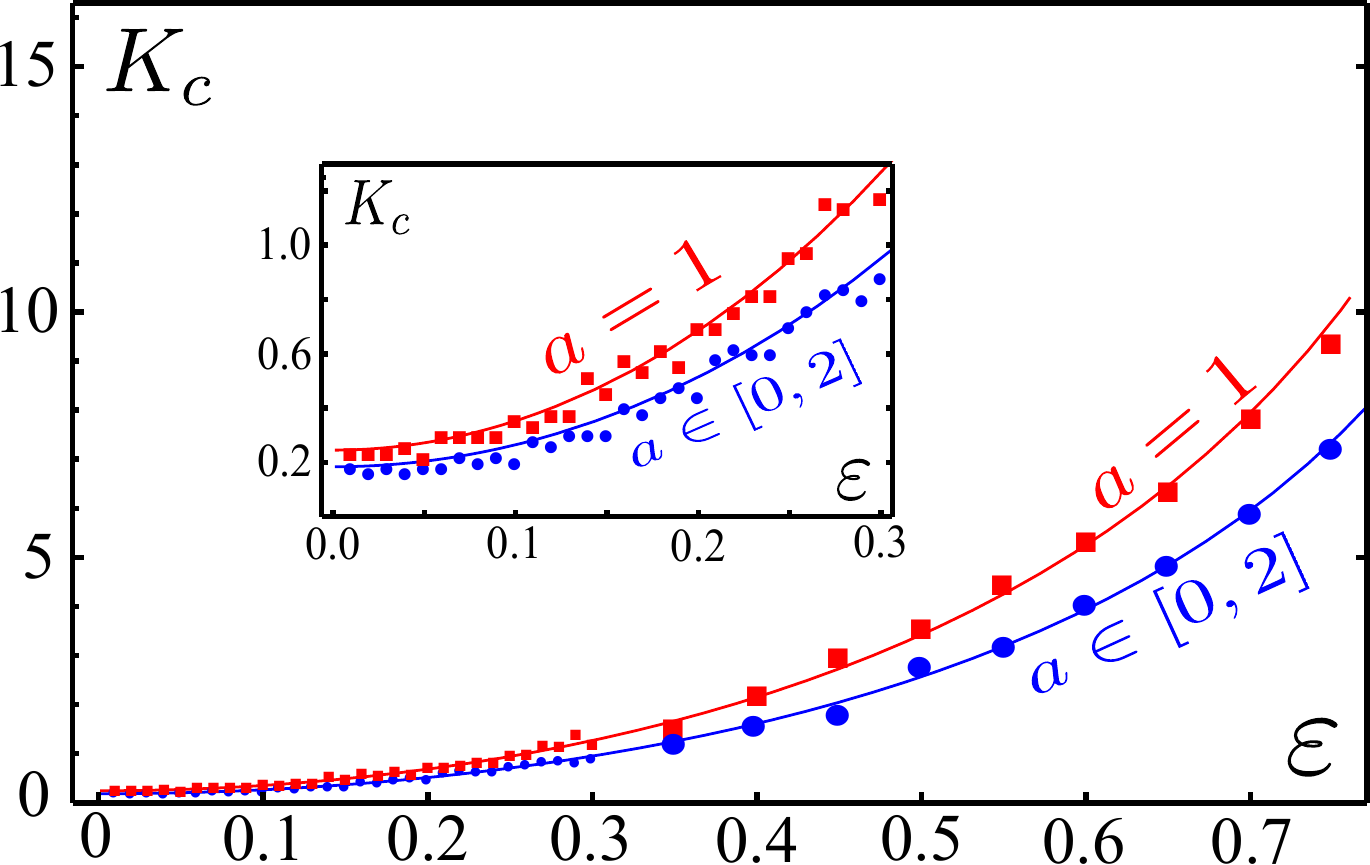}
    \caption{Estimated (symbols) and theoretical (solid line) values of $K_c$ as a function of $\varepsilon$ for example (ii). Inset: enlargement of the region $0\leq \varepsilon \leq 0.3$.}\label{colors}
\end{figure}

For example (ii) we compare the case in which $a = 1$ for all oscillators to that in which $a$ is uniformly distributed in $[0,2]$. The masses are again uniformly distributed in $[1-\varepsilon,1+\varepsilon]$. Numerical estimates of $K_c$ as a function of $\varepsilon$ are compared to the predictions (solid lines) in Fig.~\ref{colors}.  Note that $K_c$ is smaller when the charges are heterogeneous, as predicted by \Eq{kcfin}. 

\begin{figure}[thb]
  \centering
    \includegraphics[width=1\linewidth]{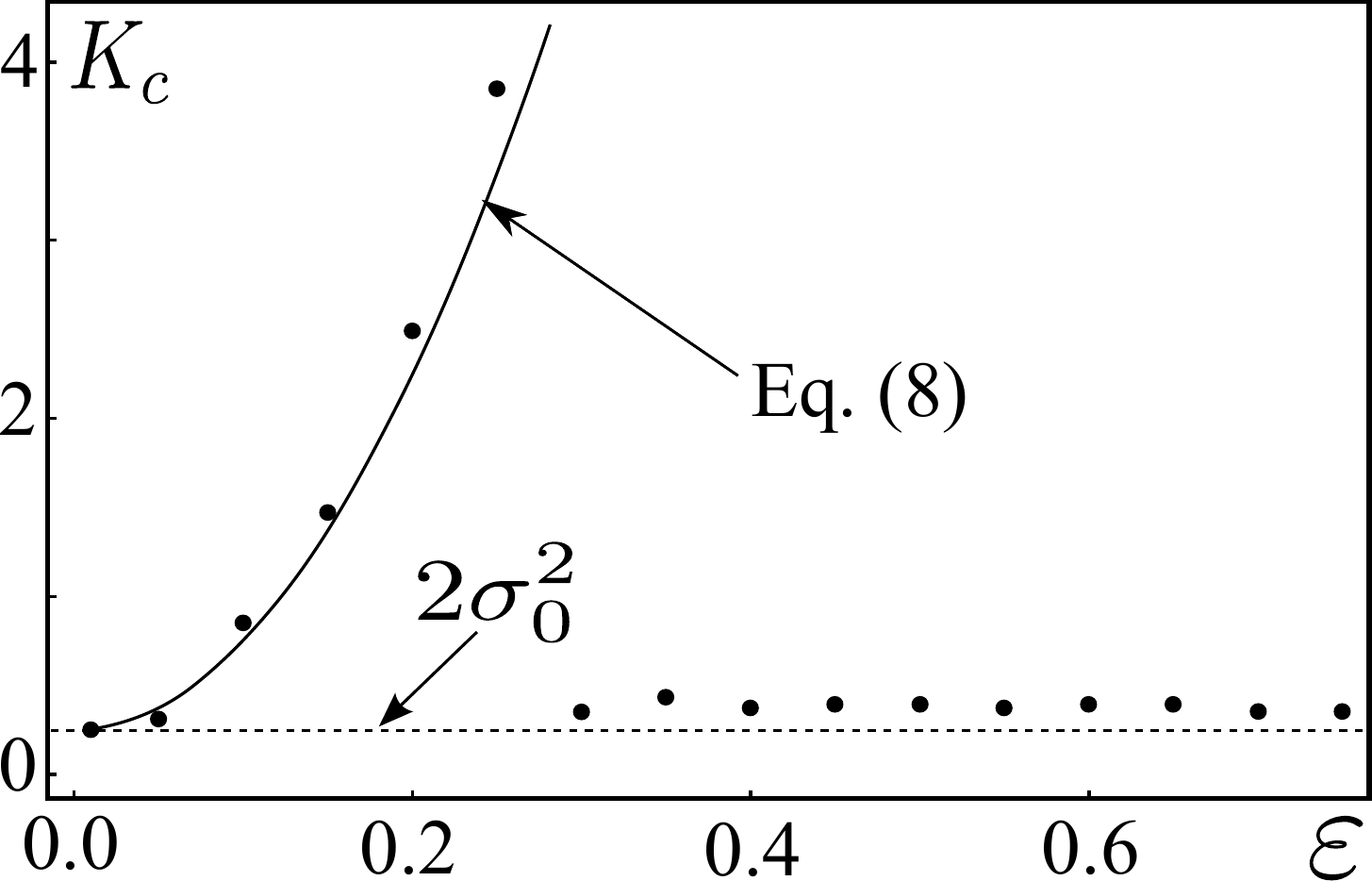}
    \caption{Simulated (symbols) and theoretical (solid line) values of $K_c$ as a function of $\varepsilon$ for example (iii). The dashed line indicates the value $2\sigma_0^2$ for $K_c$ predicted in the absence of disorder.}\label{boom}
\end{figure}

For example (iii) we show a case in which the assumption of a narrow frequency distribution is violated and our analysis breaks down. Now the distribution of masses and charges is
\[
	h(I,a) = \tfrac12 \left[ \delta(I - (1+\varepsilon)) + \delta(I - (1-\varepsilon))\right] 
	        \delta(a-1),
\]
 i.e., two peaks equidistant from $I = 1$, with the charges set to $a = 1$. The initial distribution $g_0(p)$ is a Gaussian with mean $P = 5$ and $\sigma_0 = 0.35$. Figure~\ref{boom} shows the numerically estimates (symbols) versus the predicted $K_c$ of \Eq{kcfin} (solid line) as a function of $\varepsilon$. These agree when the separation between the peaks ($2\varepsilon$) is small; however, when $\varepsilon$ becomes too large, the onset of instability suddenly drops near to the value $2\sigma_0^2$ (dashed line) predicted in the absence of disorder. In this second regime the stationary distribution $g_{I,a}(p)$ for each value of $I$ is observed to have mean $P$, instead of shifting to the predicted $I \Omega$ of \Eq{finaldist}. A more detailed study of this transition is left for future research.

\section{Conclusion}\label{conclusion}

In conclusion, we have investigated analytically and numerically the stability of the incoherent state in the disordered Hamiltonian Mean Field model. We found that finite size effects can induce correlations between the momenta, moments of inertia, and coupling constants of the rotors, modifying the onset of instability of the incoherent state. Indeed, heterogeneity in the moments of inertia tends to stabilize the incoherent state, while heterogeneity in the coupling strengths tends to destabilize the incoherent state. For sharply peaked parameter distributions, we developed an analytical formula for the modified critical coupling strength. Our analysis also qualitatively describes the behavior observed for broader distributions. Finally, we discovered a novel transition for a bimodal distribution of masses. Our results provide new insights into the factors affecting the phase transition in the  HMF model, an iconic testbed for the study of long-range Hamiltonian systems, and provide a motivation to search for analogous results in more specific systems.
\vspace{-0.7cm}
\begin{acknowledgments}
JGR acknowledges a useful discussion with Ed Ott. JDM acknowledges the support of NSF grant DMS-1211350. 
\end{acknowledgments}

\bibliography{hamilsync}

\end{document}